\documentclass[aps,showpacs,superscriptaddress,nofootinbib,floatfix,mamsmath,11pt]{revtex4}
\usepackage{epsfig,epsf,graphics,psfrag}
\usepackage{times}
\usepackage{color}
\usepackage{amsfonts,amssymb,stmaryrd,latexsym,amsmath}
\usepackage{bm}
\usepackage{enumerate}
\usepackage{enumitem}
\usepackage{hhline}

\newcommand{\be}{\begin{equation}}
\newcommand{\ee}{\end{equation}}
\newcommand{\ba}{\begin{eqnarray}}
\newcommand{\ea}{\end{eqnarray}}

\begin{document}
\title{Where is the $\bm{\chi_{c0}(2P)}$?}

\author{Feng-Kun Guo}
 \email{fkguo@hiskp.uni-bonn.de}
 \affiliation{Helmholtz-Institut f\"ur Strahlen- und
             Kernphysik and Bethe Center for Theoretical Physics, \\
             Universit\"at Bonn,  D--53115 Bonn, Germany}

\author{Ulf-G. Mei{\ss}ner}
 \email{meissner@hiskp.uni-bonn.de}
 \affiliation{Helmholtz-Institut f\"ur Strahlen- und
             Kernphysik and Bethe Center for Theoretical Physics, \\
             Universit\"at Bonn,  D--53115 Bonn, Germany}
 \affiliation{Institute for Advanced Simulation, Institut f\"{u}r Kernphysik
             and J\"ulich Center for Hadron Physics, \\
             Forschungszentrum J\"{u}lich, D--52425 J\"{u}lich, Germany}

\begin{abstract}
\noindent Although the analysis of the BaBar Collaboration prefers $J^P=0^+$ for
the $X(3915)$, it is difficult to assign the $X(3915)$ to the $\chi_{c0}(2P)$. We
show that there is an indication of the $\chi_{c0}(2P)$ with a mass around 3840~MeV
and width of about 200~MeV in the Belle and BaBar data for $\gamma\gamma\to D\bar D$.
\end{abstract}


\pacs{14.40.Pq, 13.25.Gv}

\maketitle

With the discovery of many new states in the charmonium mass region, one faces
the problem of understanding their nature (for a comprehensive review, see
Ref.~\cite{Brambilla:2010cs}). Among these $XYZ$ states, the $Z(3930)$ was
discovered by the Belle Collaboration in the process $\gamma\gamma\to D\bar
D$~\cite{Uehara:2005qd}, and later on confirmed by the BaBar Collaboration in the
same process~\cite{Aubert:2010ab}. The experimantal results on mass, angular
distributions, and $\Gamma(Z(3930)\to\gamma\gamma)\mathcal{B}(Z(3930)\to D\bar D)$
are all consistent with the expectation for the $\chi_{c2}(2P)$, which is the
radially excited $P$-wave tensor charmonium state. The mass was measured to be
$3927.2\pm2.6$~MeV~\cite{PDG}. It is below the $D^*\bar D^*$ threshold so that it
decays dominantly into the $D\bar D$ pair in a $D$-wave, and its decay width is
$24\pm6$~MeV~\cite{PDG}. So far, this is the only unambiguously identified radially
excited $P$-wave charmonium state.

In the same mass region, another structure was reported first by the Belle
Collaboration~\cite{Abe:2004zs} in the $J/\psi\omega$ invariant mass distribution
in exclusive $B\to K\omega J/\psi$ decays, and confirmed later by the BaBar
Collaboration with more statistics~\cite{Aubert:2007vj,delAmoSanchez:2010jr}. This
structure is identified as the same state observed in the same final states with
similar mass in the $\gamma\gamma\to J/\psi\omega$ process~\cite{Uehara:2009tx}. It
is called $X(3915)$, and its mass and width are $3917.5\pm2.7$~MeV and
$27\pm10$~MeV, respectively~\cite{PDG}. Very recently, a spin-parity analysis has
been performed for the process $X(3915)\to J/\psi\omega$ by the BaBar
Collaboration~\cite{Lees:2012xs}, and the results suggest that the quantum numbers
of this state are $J^P=0^+$. It was therefore identified as the $\chi_{c0}(2P)$
state~\cite{Lees:2012xs} following the suggestion of Ref.~\cite{Liu:2009fe}.
However, assigning the $X(3915)$ as the $\chi_{c0}(2P)$ state faces the
following problems:
\begin{enumerate}[label=(\arabic*),topsep=0pt,partopsep=0pt,itemsep=0pt,parsep=0pt]
\item As pointed out in, for instance,
    Refs.~\cite{Brambilla:2010cs,Eichten:2005ga}, the partial decay width of
    the $X(3915)\to J/\psi\omega$ would be $\gtrsim1$~MeV if it is produced
    similarly in $B$-meson decays as the well-understood conventional
    charmonium states. This estimate is consistent with the one using
    $\Gamma(X(3915)\to\gamma\gamma) \mathcal{B}(X(3915)\to J/\psi\omega)$,
    which was reported to be $(52\pm10\pm3)$~eV, if its spin is 0, by the BaBar
    Collaboration~\cite{Lees:2012xs}. The value is consistent with an earlier
    measurement by the Belle Collaboration~\cite{Uehara:2009tx}. Were the
    $X(3915)$ the $\chi_{c0}(2P)$ state, its width into two photons would be
    similar to that of the $\chi_{c0}$~\footnote{Assuming that the $D\bar D$
    modes dominate the decays of the $\chi_{c2}(2P)$,
    $\Gamma(\chi_{c2}(2P)\to\gamma\gamma)$ is about 0.2~keV~\cite{PDG}. It is
    of the same order as $\Gamma(\chi_{c2}\to\gamma\gamma)=0.51\pm0.04$~keV.}.
    Given $\Gamma(\chi_{c0}\to\gamma\gamma)=2.3\pm0.2$~keV, it is reasonable to
    assume $\Gamma(\chi_{c0}(2P)\to\gamma\gamma)\sim1$~keV. Identifying the
    $X(3915)$ with the $\chi_{c0}(2P)$ would give the same estimate as before,
    $\Gamma(X(3915)\to J/\psi\omega)\gtrsim1$~MeV. Such a value is at least
    one-order-of-magnitude larger than the OZI (Okubo--Zweig--Iizuka)
    suppressed hadronic widths of the $\psi(2S)$ and the $\psi(3770)$.
\item Normally, the dominant decay channels of a scalar meson should be the
    open-flavor modes, i.e. OZI allowed, if the meson lies above the
    corresponding thresholds. However, the $X(3915)$ was not observed in the
    $D\bar D$ channel, despite the facts that they can couple in an $S$-wave
    and the $D\bar D$ threshold, 3730~MeV, is about 200~MeV lower than the
    $X(3915)$ mass.~\footnote{There has been a suggestion that the $X(3915)$
    couples to the $D\bar D$~\cite{Chen:2012wy}. However, the result for the
    width obtained from their fit to the $\gamma\gamma\to D\bar D$ data of both
    the Belle~\cite{Uehara:2005qd} and Babar~\cite{Aubert:2010ab}
    Collaborations, $8.1\pm9.7$~MeV, is consistent with 0. This implies that
    the $X(3915)$ only plays a minor role in their fit.} Furthermore, when
    there is no suppression due to either isospin or SU(3) breaking, from the
    OZI rule or because of tiny phase space, all known hadrons which decay in
    an $S$-wave have widths of order 100~MeV or even more. The width of the
    $X(3915)$, being comparative to that of the $D$-wave decaying
    $\chi_{c2}(2P)$, seems too small.
\item As mentioned in Ref.~\cite{Guo:2010ak}, the mass difference between the
    $\chi_{c2}(2P)$ and the $X(3915)$, $9.7\pm3.7$~MeV, is too small for the
    fine splitting of $P$-wave charmonia. It is one-order-of-magnitude smaller
    than the fine splitting of $1P$ states
    $M_{\chi_{c2}}-M_{\chi_{c0}}=141.45\pm0.32$~MeV~\cite{PDG}. Furthermore, it
    is even smaller than the analogous splitting in bottomonium systems,
    $M_{\chi_{b2}(2P)}-M_{\chi_{b0}(2P)}=36.2\pm0.8$~MeV~\cite{PDG} even though
    the Hamiltonian terms responsible for the fine splitting are proportional
    to $1/m_Q^2$, with $m_Q$ the heavy quark mass.
\end{enumerate}

So where should the $\chi_{c0}(2P)$ be? There have been a few lattice calculations
on the mass spectrum of excited charmonium states. For the $\chi_{c0}(2P)$ state, a
quenched calculation gives a mass of $4091\pm61$~MeV~\cite{Dudek:2007wv}, while it
is predicted to be around 4~GeV in full QCD with a pion as large as
1~GeV~\cite{Bali:2011rd}. However, effects due to light sea quarks / large pion
masses could be significant for the excited $P$-wave charmonia nearby open-charm
thresholds~\cite{Guo:2012tg}. Results calculated with a lower pion mass, though
still around 400~MeV, are presented in Ref.~\cite{Liu:2012ze}. The authors
calculated the mass differences between the excited charmonia and the $\eta_c$ in
order to reduce the systematic uncertainty due to tuning the charm quark mass. The
results are $972\pm9$~MeV and $1041\pm12$~MeV for the $\chi_{c0}(2P)$ and the
$\chi_{c2}(2P)$, respectively. If the experimental mass of the $\eta_c$ is used,
the mass of the $\chi_{c0}(2P)$ would be $3953\pm9$~MeV, and meanwhile the
$\chi_{c2}(2P)$ would be about 100~MeV heavier than the measured mass. However, one
may extract the fine splitting between the $\chi_{c2}(2P)$ and the $\chi_{c0}(2P)$
from their calculation, which is $69\pm15$~MeV. Using the experimental mass of the
$\chi_{c2}(2P)$, one gets $M_{\chi_{c0}(2P)}=3858\pm15$~MeV. We regard this value
as the most reliable lattice estimate of the mass of the $\chi_{c0}(2P)$ obtained
so far. Quark model predictions are in the same region. For instance, the masses of
the $\chi_{c0}(2P)$ and the $\chi_{c2}(2P)$ are predicted to be 3916~MeV and
3979~MeV in the Godfrey-Isgur relativized quark
model~\cite{Godfrey:1985xj,Barnes:2005pb}, respectively. Shifting the
$\chi_{c2}(2P)$ to the observed mass and keeping the value of the fine splitting
would give 3856~MeV for the mass of the $\chi_{c0}(2P)$. A recent quark model
calculation using a screened potential~\cite{Li:2009zu} predicts a mass of 3842~MeV
for the $\chi_{c0}(2P)$; meanwhile, the prediction for the $\chi_{c2}(2P)$,
3937~MeV, agrees with the experimental value.

In the data of the process $\gamma\gamma\to D\bar D$ from both  Belle and
BaBar, there is a broad bump below the narrow peak of the
$\chi_{c2}(2P)$. The Belle Collaboration fits the $D\bar D$ invariant mass spectrum
in the region 3.80~GeV$<w<$4.2~GeV with a Breit-Wigner function for the
$\chi_{c2}(2P)$ plus a background function $\propto
w^{-\alpha}$~\cite{Uehara:2005qd}, with $w$ the invariant mass of the $D\bar D$
pair. The BaBar Collaboration fits the spectrum starting from the $D\bar D$
threshold with a background function $\propto\sqrt{w^2-m_t^2}(w-m_t)^\alpha
\exp[-\beta(w-m_t)]$, where the $D\bar D$ threshold is represented by
$m_t$~\cite{Aubert:2010ab}. However, if the $\chi_{c0}(2P)$ has a mass around
3850~MeV as estimated above, it would lie in the region of the broad bump, and
it could get hidden in such fits. In this paper, we try to fit both data
sets with two Breit-Wigner functions. Our results indicate that the $\chi_{c0}(2P)$
could have a mass around 3840~MeV.

\medskip

We will assume that all the cross sections of the $D\bar D$ production in
photon-photon collisions are due to resonant structures. This means that we will
neglect backgrounds due to non-resonant contribution and semi-inclusive $D\bar D X$
with undetected $X$ ($X$ can be soft pions or photons). We use the same
Breit-Wigner function for the resonances as the one used by the BaBar
Collaboration~\cite{Aubert:2010ab}. Taking into account both the phase space and
Blatt-Weisskopf centrifugal barrier factor $F_L$, the function reads as
\be%
B_L(w) = \left(\frac{p}{p_0}\right)^{2L+1} \frac{M}{w}
\frac{F_L^2(w)}{(w^2-M^2)^2+\Gamma^2(w)M^2},
\ee%
where $L$ is the orbital angular momentum of the $D\bar D$ pair,  $p$ is the
momentum at the center-of-mass (cm) frame, $p_0$ is the cm  momentum at $w=M$,
which is the Breit-Wigner mass of the resonance, and the energy-dependent width is
given by
\be%
\Gamma(w) = \Gamma \left(\frac{p}{p_0}\right)^{2L+1} \frac{M}{w} F_L^2(w),
\ee%
with $\Gamma$ being the width of the resonance at rest. The centrifugal barrier
factor~\cite{BlattWeisskopf,VonHippel:1972fg} is $F_0=1$ for an $S$ wave, and
\be%
F_2(w) = \frac{\sqrt{(R^2 p_0^2-3)^2+9R^2 p_0^2 }}{\sqrt{(R^2 p^2-3)^2+9R^2 p^2 }}
\ee%
for a $D$ wave. The same value 1.5~GeV as used in Ref.~\cite{Aubert:2010ab} will be
taken for the ``interaction radius'' $R$.

\begin{figure}[t]
\centering
\includegraphics[width=0.49\linewidth]{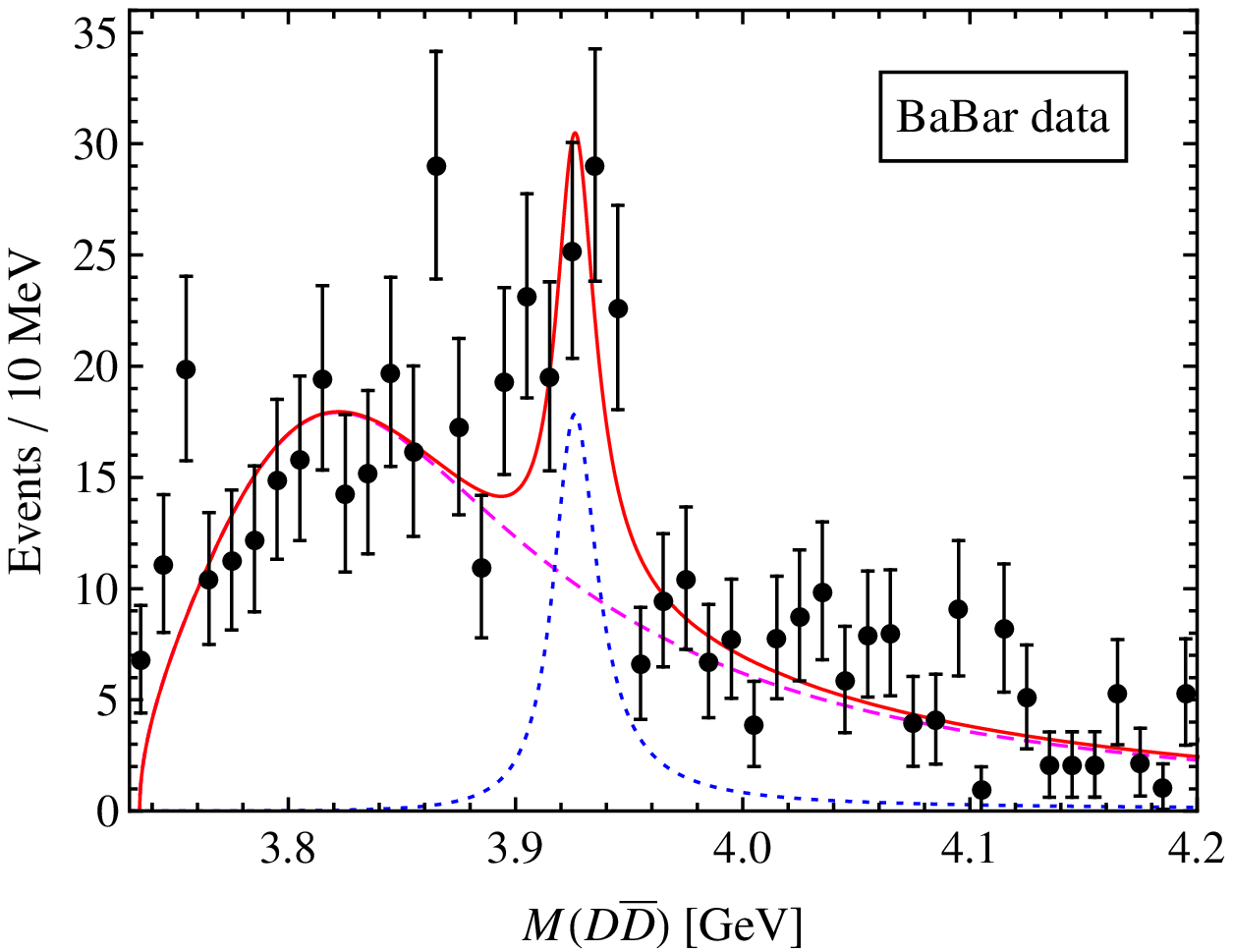}
\hfill\includegraphics[width=0.49\linewidth]{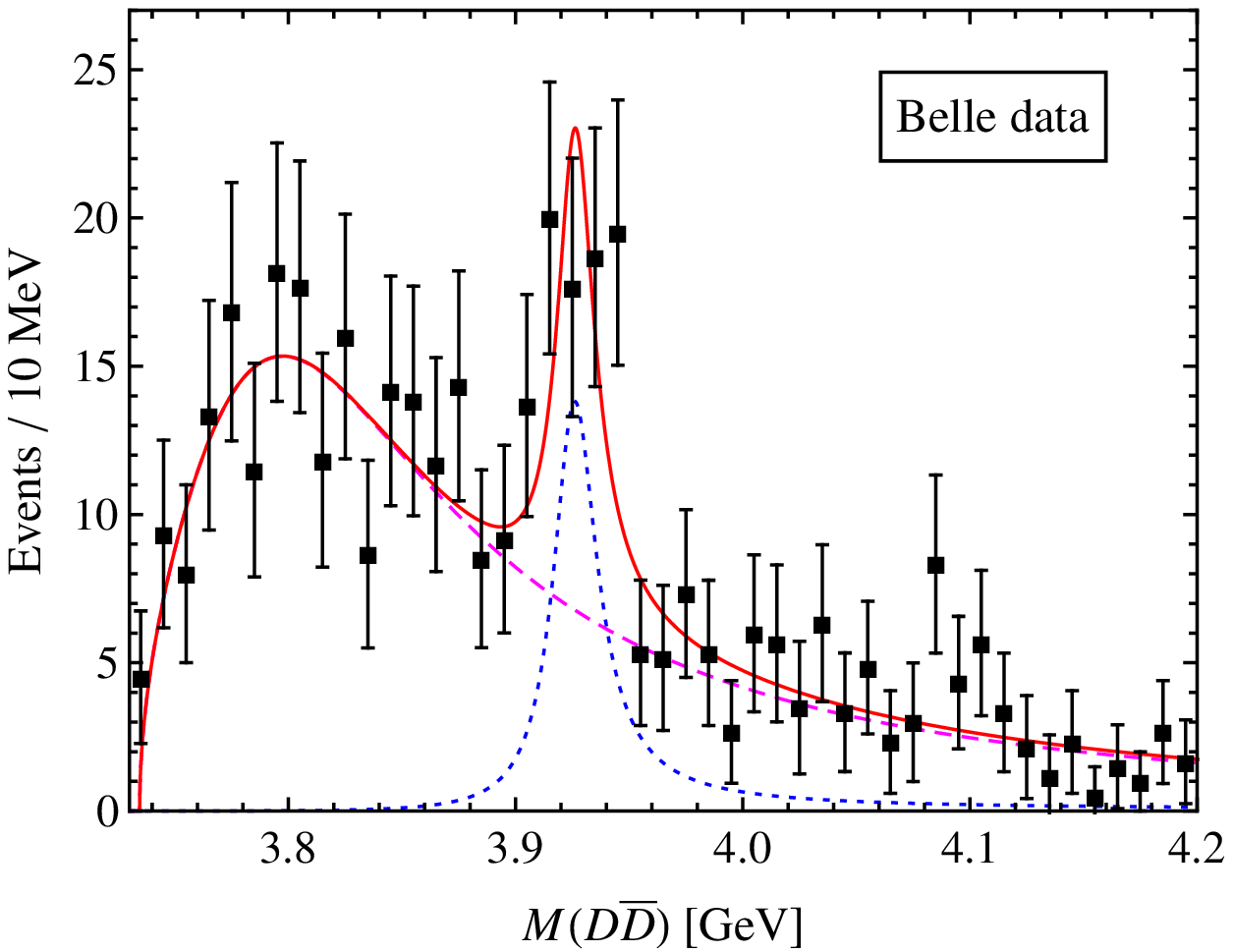}
\caption{ Fit to the BaBar (left) and Belle (right) data separately. The $D$-mass sidebands have
been subtracted from the Belle data. The dashed and dotted lines represent the
contributions from the $\chi_{c0}(2P)$ and the $\chi_{c2}(2P)$, respectively.
\label{fig:fit}}
\end{figure}

We fit to the BaBar  and the Belle data separately in the region from the
$D\bar D$ threshold to 4.2~GeV with four parameters: the mass $M_0$ and width
$\Gamma_0$ of a $0^+$ resonance which couples to the $D\bar D$ in an $S$-wave, and
two normalization constants $N_0$ and $N_2$ for the scalar meson and the
$\chi_{c2}(2P)$, respectively. The mass and width of the $\chi_{c2}(2P)$ are fixed
to 3927~MeV and 24~MeV~\cite{PDG}, respectively. There is no interference between
these two structures because they are in different partial waves. Contrary to the
BaBar data, the Belle data are not efficiency corrected. Nevertheless, the Belle
efficiency only decreases by 10\% for an increase of the invariant mass from 3.8 to
4.2~GeV, and there is no fine structure in the efficiency and background
distributions~\cite{Uehara:2005qd,Uehara}. Furthermore, the $D$-mass sidebands have
been subtracted from the Belle data used in our fit. A comparison of the best fit
to the data are shown in Fig.~\ref{fig:fit}.

The fit results are collected in Table~\ref{tab:results}, where the
uncertainties only reflect the statistical errors in the fit.
\begin{table}
\begin{ruledtabular}
\begin{tabular}{|l | c c c c |}
Data sets & $\chi^2/{\rm dof}$ & $M_0$~(MeV) & $\Gamma_0$~(MeV) & $N_2/N_0$ \\
\hline
 BaBar~\cite{Aubert:2010ab}  & 1.63  & $3848.1\pm7.7$ & $229\pm26$ & $0.0121\pm0.0034$\\
 Belle~\cite{Uehara:2005qd}  & 0.79  & $3825.1\pm8.4$ & $212\pm29$ & $0.0132\pm0.0046$\\
\end{tabular}
\end{ruledtabular}
\caption{\label{tab:results} Results of fitting to the BaBar and Belle data,
respectively. See text for details.}
\end{table}
One sees that the two resonance assumption gives a reasonable fit to both data sets.
The large value of $\chi^2/{\rm dof}$ for the fit to the BaBar data comes mainly
from a few bins where the event numbers are quite separated from their
neighbors. Comparing the resulting parameters from the two fits, the difference in
the values of the mass is $2\sigma$, and the values of the width and the ratio of
the normalization constants are fully consistent with each other. The mass is
compatible with the lattice estimate for the mass of the $\chi_{c0}(2P)$ discussed
above, and the width is of the right order for an $S$-wave strongly decaying
hadron. Furthermore, its main decay channel should be $D\bar D$ since such a
structure has never been reported elsewhere. All these properties indicate that the
broad bump could be assigned as the $\chi_{c0}(2P)$. The weighted average~\cite{PDG} of
both fits gives
\be%
 M_{\chi_{c0}(2P)} = (3837.6\pm11.5)~{\rm MeV}, \qquad
 \Gamma_{\chi_{c0}(2P)} = (221\pm19)~{\rm MeV}.
\ee%

In summary, it is difficult to accommodate the $X(3915)$ to the still missing
$\chi_{c0}(2P)$. Being a $J^P=0^+$ particle, the $X(3915)$ is probably of exotic
nature. There is an indication that the present data of the $\gamma\gamma\to D\bar
D$ process already contain signals of the $\chi_{c0}(2P)$ with a mass and width
around 3840~MeV and 200~MeV, respectively. More refined analysis of the data with
higher statistics is definitely necessary to confirm our assertion. In addition to
the photon-photon fusion process, searching for the $\chi_{c0}(2P)$ in the $D\bar
D$ distribution can also be done at BES-III using radiative decays of higher vector
charmonia~\cite{Li:2012vc}, and at the hadron colliders LHC and FAIR.

\section*{Acknowledgments}
We would like to thank Sadaharu Uehara and Torsten Schr\"oder for useful
communications. This work is supported in part by the DFG and the NSFC through
funds provided to the Sino-German CRC 110 ``Symmetries and the Emergence of
Structure in QCD'' and the EU I3HP ``Study of Strongly Interacting Matter'' under
the Seventh Framework Program of the EU. U.-G. M. also thanks the BMBF for support
(Grant No. 06BN7008). F.-K. G. acknowledges partial support from the NSFC (Grant
No. 11165005).

\end{document}